\documentclass{ifacconf}                                  
\usepackage{cite}
\usepackage{natbib}        
\usepackage{amsmath,amssymb,amsfonts}
\usepackage{algorithmic}
\usepackage{graphicx}
\usepackage{textcomp}
\usepackage{xcolor}
\usepackage{breqn}
\usepackage{booktabs}
\usepackage{subcaption}
\usepackage{fancyhdr}

\newcommand{\cM}{{\mathcal M}}

\newtheorem{definition}{Definition}
\newtheorem{assumption}{Assumption}

\begin{document}

 \thispagestyle{fancy}
\begin{frontmatter}
Published in 2021 Measurements Estimation and Controls Conference, Oct 26, 2021

\title{Information-Theoretic Approach for Model Reduction Over Finite Time Horizon 
}

\author[First]{Punit Tulpule}
\author[Second]{Umesh Vaidya}

\address[First]{Mechanical and Aerospace Engineering Department, The Ohio State University, Columbus OH 43210\\ (e-mail: tulpule.3@ osu.edu)}

\address[Second]{Department of Mechanical Engineering, Clemson University, Clemson SC 29634\\ (e-mail: uvaidya@clemson.edu)}

\begin{abstract}
This paper presents an information-theoretic approach for model reduction for finite time simulation. Although system models are typically used for simulation over a finite time, most of the metrics (and pseudo-metrics) used for model accuracy assessment consider asymptotic behavior e.g., Hankel singular values and Kullback-Leibler(KL)  rate metric. These metrics could further be used for model order reduction. Hence, in this paper, we propose a generalization of KL divergence-based metric called n-step KL rate metric, which could be used to compare models over a finite time horizon.  We then demonstrate that the asymptotic metrics for comparing dynamical systems may not accurately assess the model prediction uncertainties over a finite time horizon.  Motivated by this finite time analysis, we propose a new pragmatic approach to compute the influence of a subset of states on a combination of states called information transfer (IT). Model reduction typically involves the removal or truncation of states. IT combines the concepts from the n-step KL rate metric and model reduction. Finally, we demonstrate the application of information transfer for model reduction. Although the analysis and definitions presented in this paper assume linear systems, they can be extended for nonlinear systems. 
\end{abstract}

\end{frontmatter}

\section{Introduction}
Model simplification, approximation, and reduction is a critical phase in the systems design process. According to \cite{gugercin2004survey}, in model reduction, the goal is to find a lower order model that has similar response characteristics as the original model. Often, the motivation behind model reduction is to improve computation time, simplify the control design process, and develop model-based control algorithms. Although different methods exist for model simplifications, the underlying philosophy relies on the concept of states' removal (or truncation). Hence, this paper is primarily focused on model order reduction methods by the removal of states. The idea behind model reduction is to first approximate the model and then compare its performance with respect to the higher-order model. Hence, to reduce the model, there is a need for a measure to compare dynamical systems. The most commonly used method for model reduction is balanced truncation. In this method, the model is first transformed such that the states which are difficult to reach are also difficult to observe.  The difficult to reach and observe states are then truncated from the model. Another approach is based on information theory. The idea is to compare the predictive uncertainties of different approximate models with the original model.  To compare the accuracy of models, \cite{georgiou2007distances} proposed an approach based on a distance between spectral densities in terms of prediction errors. Motivated by different classical distances (and pseudo distances) \cite{6165647} proposed spectral density estimation approaches and established a connection between time and spectral domain-relative entropies. Based on Georgiou's work, \cite{deng2012model,yu2009kullback, yu2010kullback} proposed a pseudo metric called Kullback Leibler (KL) rate metric to compare dynamical models. KL rate metric has been used for comparing dynamical models in various applications, for example, \cite{DENG20141188, deng2011optimal}. The KL-rate pseudo metric was further extended for Markov chain model reduction by \cite{deng2012model},  but it relies on the asymptotic computation of probability density functions (pdf). In other words, the KL-rate pseudo metric can only compare models at a steady-state, but in the case of dynamical systems, the modeling accuracy is also essential during transience. Consider a system with two-time constants, one very slow and the other very fast, for example, a battery charge-discharge model augmented by its aging dynamic. Removing the fast dynamic could result in significant errors on small time scales but may have small errors in predicting or estimating the age of the battery.  This paper first demonstrates that the asymptotic assessment comparison between models may not correctly indicate the accuracy over finite time simulation.  This paper proposes an extension to the KL rate metric called n-step KL rate pseudo metric to compare approximate models generated by removing states over a finite time horizon. 
When it comes to model reduction, we need to compare models with a different number of states. The process of model reduction motivated \cite{san2005information} to utilize further the concept of 'freezing of states' along with the definition of n-step KL rate metric. Hence, we propose a new definition of Information Transfer (IT). IT could be employed to decide which states to truncate with the least increase in model uncertainty/error. 
IT is the entropy transferred from a subspace of states to the observables. In most cases, the removal of states may increase the uncertainty in the predictions made by a model.  IT is essentially the change in the uncertainty in the predictions of observables due to the 'freezing' of states as proposed by \cite{san2005information} is the concept of freezing part of the system dynamics and has similarity with the definition of IT proposed by \cite{sinha2016causality,sinha2017information}. The idea of freezing alleviates the problem associated with other information-based causality measures and captures the direct causal links as shown by \cite{sinha2016causality}. However, unlike the previous research by \cite{sinha2020data}, in the proposed definition of IT, the 'change in uncertainty' is computed using KL divergence. 
In summary, the essential novelty of the proposed definition of IT is that it quantifies the influence of a subset of states on other states and a combination of states. Unlike the n-step KL rate metric, which is a metric to compare two models, IT is a property of a system or a model of a system. Also, the new definition allows the evaluation of dynamical systems over a finite time horizon. The paper purposefully uses simple examples to better understand the concept, but the proposed definition could be further applied to multi-variate and non-Gaussian uncertainty models. 

The paper is organized as follows. In the second section of the paper, we discuss model reduction using asymptotic analysis. In the third section, we demonstrate using a simple example that comparison based on asymptotic may not provide correct assessment for finite time horizon applications.  A new definition of IT which combines the concepts of n-step KL rate metric and truncation of states, is proposed in section IV. In this section, we also demonstrate the use of IT for model reduction. Finally, in the last section, we summarize the conclusions drawn from the work.

\section{Model Reduction By Asymptotic Comparison}
{
The model reduction process has two steps, namely 1) generate a set of simplified models by removal or truncation of states and 2) compare the simplified models with the original model using some metric. An additional balancing step may be performed before model reduction. Removal of the state is essentially considering that the states are stationary. The second step reveals the best-reduced model which is dependent on the choice of metric. 
Consider a linear system $\cM0$ shown below.  
\begin{eqnarray}
\cM0: 
\begin{cases} 
\overbrace{\begin{bmatrix} a_{t+1} \\ b_{t+1} \end{bmatrix}}^{x_{t+1}} = \overbrace{\begin{bmatrix} A_{11} & A_{12} \\ A_{21} & A_{22}\end{bmatrix}}^{A_0} \overbrace{\begin{bmatrix} a_t \\ b_t \end{bmatrix}}^{x_t} + \begin{bmatrix} B_1 \\ B_2 \end{bmatrix} d_t \\
y_t = \overbrace{\begin{bmatrix} C_1 & C_2 \end{bmatrix}}^C x_t + Dd_t
\end{cases}
\label{model1}
\end{eqnarray}


where $x\in \Re^m$ is a state vector, $y\in\Re$ is output, $A$, $B$, $C$ are real matrices such that $A$ is Piccard (has all eigenvalues inside the unit circle), the pair $(A,C)$ is observable. In this paper we are considering single output for the sake of simplicity of notation. The input $d_t\sim\mathcal{N}(0,1)$ is i.i.d Gaussian random variable. The state space could be divided into two subspaces according to $a\in \Re^{m_1}$ and $b\in \Re^{m_2}$ such that $m_1 + m_2 = m$. 
Let's assume that a reduced representation of $\cM0$ could be obtained by considering that states $a$ are constant i.e. $a=\bar{a}$. Hence, the dynamics of the approximate model is $b_{t+1}=A_{22}b_{t}+A_{21}\bar{a}+B_2d_t$.  To simplify the analysis, we will equivalently write the simplified model as

\begin{eqnarray}
\cM1: 
\begin{cases} 
w_{t+1} = A_1w_t + Bd_t \\
z_t = C w_t + Dd_t
\end{cases}
\end{eqnarray}
where 
\[ 
A_1 = \begin{bmatrix} 1 & 0 \\ A_{21} & A_{22}\end{bmatrix} 
\]
Also, $w^\top = \begin{bmatrix} a^\top & b^\top \end{bmatrix} $, $a_0 = \bar{a}$ is a constant, $w\in\Re^m $ is state vector, and $z\in \Re$ is the output of $\cM1$. Note that $B_1$ is not set to zero to emphasize that uncertainty in states $a$ still persists in the simplified model. To quantify the error between model $\cM1$ and $\cM0$, different metrics could be used.  For example, balanced truncation uses the steady-state observability and controllability to obtain balanced state space and then use Hankel singular values to truncate states. Similarly, the KL rate metric relies on asymptotic KL divergence between prediction uncertainty. 

\subsection{Balanced Truncation} 
Consider a linear system $\cM0$. The controllability and observability gramians ($P$ and $Q$) are the positive definite solutions to following asymptotic Lyapunov equations
\[ A_0P+PA_0^T=-BB^\top\] and
\[A_0^\top Q+QA_0 = C^\top C\]
The states for which the Hankel singular values are small are truncated. The loss in accuracy due to removal of states in terms of $H_\infty$ norm is obtained by removing corresponding singular values. Note here that the balancing as well as the Hankel singular values essentially rely on asymptotic behavior.
}

\subsection{KL Rate Metric}
Prashant Mehta et al. proposed a pseudo metric called Kullback-Leibler rate metric to compare stochastic dynamical system models \cite{deng2012model}. 

Consider the two linear dynamical systems $M_i$ for $i=0,1$. We know that $\cM0$ is the more accurate or the 'true' model and $\cM1$ is a approximate model. 
Let $p(x_t)$ and $q(w_t)$ be the probability density functions of states at time $t>0$. Similarly, let $p(y_t)$ and $q(z_t)$ be the probability density functions of measurement variables. 
\newline \textbf{Notation:} 
\begin{itemize}
    \item For any random variable $x(t)$, the time history of the variable from time $t = n$ to $t = k>n$ is written as $x_n^m$ i.e. $x_n^k = \begin{bmatrix}x_n, & x_{n+1}, & \cdots, & x_k \end{bmatrix}$
    \item The probability of random variable $x_0^n$ is simply written as $p(x_0^n)=p_n(x)$
\end{itemize}
The propagation of probability density function under system dynamics \ref{model1} could be considered as a belief process. The propagation of belief $p$ from time $t$ to $t+1$ is obtained by considering the filtering problem involving two terms - the prior and the evidence \cite{chen2003bayesian} as shown in equation below.
\begin{eqnarray}
p(x_{t+1}|y_0^t) &=& \int_X p(x_{t+1}|x_t)p(x_t|y_0^t)dx_t \nonumber \\
p(y_{t+1}|y_0^t) &=& \int_X p(y_{t+1}|x_{t+1})p(x_{t+1}|y_0^t) dx_t
\label{Eq:filter1}
\end{eqnarray}
A similar belief process can be constructed for Model $\cM1$.

\begin{eqnarray}
q(w_{t+1}|y_0^t) &=& \int_X q(w_{t+1}|w_t)q(w_t|y_0^t)dw_t \nonumber \\
q(z_{t+1}|y_0^t) &=& \int_X q(z_{t+1}|w_{t+1})q(w_{t+1}|y_0^t) dw_t
\label{Eq:filter2}
\end{eqnarray}
Note here that the belief process for model $\cM1$ uses the time-history of measurements from the 'truth' model $\cM0$.
The KL rate pseudo-metric is then defined as \cite{deng2012model}
\begin{equation}
\Delta \mathcal{H}(\cM0,\cM1) = \lim_{t\rightarrow \infty }E_{p_t(y)}\left(\ln{\frac{p(y_t|y_0^{t-1})}{q(z_t|y_0^{t-1})}} \right)
\label{Eq:KLRateDef}
\end{equation}
where $E_{p_t(y)}$ is the expectation taken over the probability of random variable $y_0^t$. For the linear Gaussian problem, the evolution of the two belief processes for models $\cM0$ and $\cM1$ are given by the Kalman filtering equations.
The Kalman filter equations for model $\cM0$ for the a priori estimate are
\begin{eqnarray}
    \hat{x}^-_{t+1} &=& A_0(I-K_tC)\hat{x}^-_{t} + A_0K_ty_t \\
    P_{t+1}^- &=& A_0P_t^-A_0^\top \nonumber \\ 
              & & - A_0P_t^-C^\top(CP_t^-C^\top + BB^\top)^{-1}CP_t^-A_0^{\top} \nonumber
\end{eqnarray}where $P_t^-$ is the a priori estimate of covariance of the state estimation error, $\hat{x}^-$ is the a priori state estimate and $K_t$ is the Kalman gain.
 Similarly for model $\cM1$
\begin{eqnarray}
    \hat{w}^-_{t+1} &=& A_1(I-K_tC)\hat{w}^-_{t} + A_1K_ty_t \\
    Q_{t+1}^- &=& A_1Q_t^-A_2^\top \nonumber \\ 
              & & - A_1Q_t^-C^\top(CQ_k^-C^\top + BB^\top)^{-1}CQ_t^-A_1^{\top} \nonumber
\end{eqnarray}
where $Q_t^-$ is the a priori estimate of covariance of the state estimation error, $\hat{w}^-$ is the a priori estimate, and $K_t$ is the Kalman gain for model $\cM1$. Also, we assume $P_0 = Q_0$, meaning the state covariance at the initial condition is the same for both the models.  Asymptotically, the two Kalman filter equations converge to Algebraic Riccati equations. For example, for Model $\cM0$, the state covariance matrix $P$ asymptotically converges to the solution of Riccati equation
\begin{equation}
P = \tilde{A}_1\left( P - \frac{PC^\top CP}{D^2+ CPC^\top} \right)\tilde{A}_1^\top
\label{Eq:SSRiccati}
\end{equation}
where $\tilde{A}_1 = (A_1 - BCD^{-1})$. Asymptotically, the belief process can be denoted as
\begin{eqnarray}
\lim_{n\rightarrow \infty}{ p_n} &=& \lim_{n\rightarrow \infty} {p(y_n|y_0^{n-1})} \sim \mathcal{N}(\hat{y},\sigma_{1}^2) \nonumber \\
\lim_{n\rightarrow \infty}{q_n} &=& \lim_{n\rightarrow \infty}{q(z_n|y_0^{n-1})} \sim \mathcal{N}(\hat{z},\sigma_{2}^2)
\label{Eq:LinSys_CondProb}
\end{eqnarray}
where $\hat{y}=C\hat{x}^-$ and $\hat{z}$ are the expected values of outputs $y$ and $z$ at steady state, and $\sigma_{1}^2 = CPC^\top+D^2$ and $\sigma_{2}^2 = CQC^\top + D^2$. Here, $P$ is the solution of Riccati equation ~(\ref{Eq:SSRiccati}), and $Q$ is a solution of a similar Riccati equation for Model $\cM1$. 
The KL rate metric given in equation ~(\ref{Eq:KLRateDef}) could be computed using the conditional probability distributions in equation ~(\ref{Eq:LinSys_CondProb}). The details of the asymptotic computation of the KL rate metric could be found in \cite{deng2012model}. 

In the next section, we propose an extension of the KL rate metric called n-Step KL rate metric. We also demonstrate that the asymptotic KL rate metric may be an incorrect assessment of models over finite time simulations. 

\section{Model Comparison Over Finite Time}
We will now slightly modify the definition of KL rate metric presented in equation ~(\ref{Eq:KLRateDef}) and call it 'n-step KL rate pseudo metric'. 
\begin{equation}
    \Delta \mathcal{H}_n(M1,M2) = E_{p_n(y)}\left(\ln{\frac{p(y_n|y_0^{n-1})}{q(z_n|y_0^{n-1})}} \right)
\end{equation}
The difference between KL rate metric and n-step KL rate metric is the time horizon. In the case of KL rate metric, the pdf are computed asymptotically whereas in the case of n-step KL rate metric, the pdf are computed over finite time. The $\Delta \mathcal{H}_\infty $ is a special case of $\Delta \mathcal{H}_n $ since $\Delta \mathcal{H}_\infty = \lim_{n\rightarrow \infty}{\Delta \mathcal{H}_n }$. In this section we illustrate the need for a finite-time metric using a simple example. For the sake of simplicity of notation we will consider a second order system 

\begin{eqnarray}
M0: \begin{cases}
    \begin{bmatrix} a_{t+1} \\ b_{t+1} \end{bmatrix} = \overbrace{\begin{bmatrix} A_{11} & 0 \\ 0 & A_{22}\end{bmatrix}}^{A_0} \begin{bmatrix} a_t \\ b_t \end{bmatrix} + Bd_t  \\
    y_t = C\begin{bmatrix} a_t & b_t\end{bmatrix}^\top
    \end{cases}
\end{eqnarray}
where, matrix $A_0$ is Piccard, $B = \begin{bmatrix} 1 & 1 \end{bmatrix}^\top$, $C= \begin{bmatrix} C_1& C_2\end{bmatrix} $, and $d_t \sim \mathcal{N}(0,\sigma^2)$ is a i.i.d. random variable with Gaussian distribution. It can be easily seen that the pair $(A,C)$ is observable if $C_1,C_2 \neq 0 $, and $(A,B)$ is controllable. Note that the difference between $\cM0$ and $M0$ is that the former is a $m^{th}$ order system, whereas the later is a second order system. 
One option to approximate this model is by considering that state $a$ is constant. Let's call the simplified linear model as $M1$.
\begin{eqnarray}
M1: \begin{cases}
   \begin{bmatrix} a_{t+1} \\ b_{t+1} \end{bmatrix} = \overbrace{\begin{bmatrix} 1 & 0 \\ 0 & A_{22}\end{bmatrix}}^A_1 \begin{bmatrix} a_t \\ b_t \end{bmatrix} + 
   B d_t  \\
    y_t = C\begin{bmatrix} a_t & b_t\end{bmatrix}^\top
    \end{cases}
\end{eqnarray}
The other option to approximate $M0$ is by considering that state $b$ is constant. Let's call this simplified model as $M2$.
\begin{eqnarray}
M2 \begin{cases} 
    \begin{bmatrix} a_{t+1} \\ b_{t+1} \end{bmatrix} =  \begin{bmatrix} A_{11} & 0 \\ 0 & 1\end{bmatrix} \begin{bmatrix} a_t \\ b_t \end{bmatrix} + Bd_t \\
    y_t = C\begin{bmatrix} a_t & b_t\end{bmatrix}^\top
    \end{cases}
\end{eqnarray}

The goal of this example is to show that deciding which approximation ($M1$ or $M2$) is better, depends on the duration of simulation. The decision is made based on KL rate metric computed between models. First we will compute the n-step KL rate metric between models $M0$ and $M1$.

\begin{thm}
For linear models $M0$ and $M1$, the n-step KL rate metric between the two is give by
\begin{equation}
    \Delta \mathcal{H}_n = \frac{C_1^2(1-A_{11})^2\Sigma_n^{11}}{2(C_1+C_2)^2\sigma^2}
    \label{Eq:nstepKLFormula}
\end{equation}
where $\Sigma_n^{11}$ is the first component of auto-covariance of states given by 
\begin{equation*}
\Sigma^{11}_{t+1} = A_{11}^2\Sigma^{11}_{t} + \sigma^2
\label{Eq:Lyap_1state}
\end{equation*}

Further, the n-step KL rate metric could be written as 
\[
    \Delta \mathcal{H}_n = (A_{11})^2\Delta \mathcal{H}_{n-1}+(1-A_{11}^2)\Delta\mathcal{H}_\infty
\]
where $\Delta\mathcal{H}_\infty$ is the asymptotic KL rate metric and $\Delta\mathcal{H}_0$ is given by 
\[ 
\Delta\mathcal{H}_0 = \frac{C_1^2(1-A_{11})^2\Sigma_0^{11}}{2(C_1+C_2)^2\sigma^2}
\].

\end{thm}

\begin{pf}
Since, the system $M0$ is observable, the knowledge of past $n$ observations $y_0^{n-1}$ are sufficient to determine the state $x_{n-1}$, for $n>2$. In such a case, the belief processes for models $M0$ and $M1$ are given by 
\[ 
p_{n}(y) = \frac{1}{\sqrt{2\pi(CB)^2\sigma^2}} exp\left( -\frac{(y-CA_0x_{n-1})^2}{2(CB\sigma)^2}\right)
\]

\[ 
q_{n}(y) = \frac{1}{\sqrt{2\pi(CB)^2\sigma^2}} exp\left( -\frac{(y-CA_1w_{n-1})^2}{2(CB\sigma)^2}\right)
\]
Note here that $2(CB)^2 = 2(C_1+C_2)^2$. Now, following example 3.2  in \cite{yu2009kullback}, it can be shown that
\[
\Delta\mathcal{H}_n(M0,M1) = \frac{C(A_1-A_0)\Sigma_n(A_1-A_0)C^\top}{2(C_1+C_2)^2\sigma^2}
\]
where, $\Sigma_n$ is the auto-covariance matrix of states $x$ computed using n-step time history of the states. 
Since the system is decoupled, it can be assumed that $\Sigma$ is block diagonal
\[ \Sigma_n = \begin{bmatrix} \Sigma_{n}^{11} & 0 \\ 0 & \Sigma_{n}^{22} \end{bmatrix}
\]
Hence,
\[ \Delta\mathcal{H}_n = \frac{C_1^2(1-A_{11})^2\Sigma^{11}_n}{2(C_1+C_2)^2\sigma^2}
\]
Since the sensor noise is not considered in this example for simplicity, the covariance matrix evolves according to the discrete time Lyapunov equation 
\begin{equation}
\Sigma_{t+1} = A_0^2\Sigma_{t} + \sigma^2
\label{Eq:DiscreteLyapunov}
\end{equation}
Since, $\Sigma$ is block-diagonal, and states $a$ and $b$ are decoupled, 
\[
\Sigma^{11}_{t+1} = A_{11}^2\Sigma^{11}_{t} + \sigma^2
\]

This proves the first part of the theorem.
The $n^{th}$ step solution of the Lyapunov equation could be written in explicit form as 
\begin{align*}
    \Sigma^{11}_n & = (A_{11})^{2n}\Sigma^{11}_0 +\sigma^2 \sum_{i=0}^{n-1} {A_{11}}^{2i} \\
    &=(A_{11})^{2n}\Sigma^{11}_0+ \frac{(1-(A_{11})^{2n})\sigma^2}{1-(A_{11})^{2}}\\
    & =
(A_{11})^{2n}\Sigma^{11}_0+(1-(A_{11})^{2n})\Sigma^{11}_{ss} 
\end{align*}

\noindent where, \[\Sigma_{ss}^{11}=A_{11}^2\Sigma_{ss}^{11}+\sigma\implies \Sigma_{ss}^{11}=\frac{\sigma}{1-A_{11}^2} \]
is the asymptotic solution of the Lyapunov equation.
Substituting the $n^{th}$ step solution in the formula for n-step KL rate metric in equation ~(\ref{Eq:nstepKLFormula})
\begin{equation}
\Delta\mathcal{H}_n = (A_{11})^{2n}\Delta\mathcal{H}_0 + (1-(A_{11})^{2n})\Delta\mathcal{H}_\infty
\label{Eq:nStepKLDynSoln}
\end{equation}
where, 
\[ 
\Delta\mathcal{H}_0 = \frac{C_1^2(1-A_{11})^2\Sigma_0^{11}}{2(C_1+C_2)^2\sigma^2}
\]
and 
\[ 
\Delta\mathcal{H}_\infty = \frac{C_1^2(1-A_{11})^2\Sigma_{ss}^{11}}{2(C_1+C_2)^2\sigma^2}
\]

The $\Delta\mathcal{H}_n$ expressed in equation ~(\ref{Eq:nStepKLDynSoln}) is a solution of the dynamical system
\begin{equation} \label{Eq:nStepKLSoln}
    \Delta \mathcal{H}_n = (A_{11})^2\Delta \mathcal{H}_{n-1}+(1-A_{11}^2)\Delta\mathcal{H}_\infty
\end{equation}

\end{pf}

The n-step KL rate metric can be viewed as a first-order dynamical system. Hence, the values of the n-step KL rate metric are determined by 1) initial condition of the dynamical system 2) time constant and 3) the steady-state value. The initial value is dependent on the system parameters $A_{11}$ and $C_1$, and an initial guess of the covariance matrix $\Sigma^{11}_0$. The time-constant is $A_{11}^2$, and steady-state value is given by the asymptotic KL rate metric.  The key point to be noted here is that the prediction uncertainty due to modeling approximation as quantified using KL divergence changes with time and asymptotically stabilizes.

To find which one of the approximate models $M1$ and $M2$ is a better approximation, the KL rate metric could be used. The higher the KL rate metric, the higher is the uncertainty in the approximate model. Hence, a model with a lower KL rate metric is considered as a more accurate approximation \cite{deng2012model}. As shown in theorem 1, the KL-rate metric varies with time. Hence, the accuracy of the approximate model changes with time. Now we will show that choosing the approximate model purely based on asymptotic metrics may lead to incorrect assumptions over shorter time-time scales.

\renewcommand{\arraystretch}{2}
\begin{table}[hb]
\begin{center}
 \captionsetup{width=\linewidth}
\caption{n-Step KL rate metric to compare models $M1$ and $M2$ with $M0$} \label{Tab:info_Tranf_formula}
\begin{tabular}{lr} \toprule
$\alpha_n = \Delta\mathcal{H}_n(M0,M1)$& \(\dfrac{C_1^2(1-A_{11})^2\Sigma_n^{11}}{2(C_1+C_2)^2\sigma^2}\) \\ 
& $\Sigma^{11}_{t+1} = A_{11}^2\Sigma^{11}_{t} + \sigma^2$ \\ \midrule
$\beta_n = \Delta\mathcal{H}_n(M0,M2)$& $\dfrac{C_2^2(1-A_{22})^2\Sigma_n^{22}}{2(C_1+C_2)^2\sigma^2}$ \\
& $\Sigma^{22}_{t+1} = A_{22}^2\Sigma^{22}_{t} + \sigma^2$ \\ \bottomrule
\end{tabular}
\end{center}
\end{table}
\renewcommand{\arraystretch}{1}

We can derive n-step KL rate metric between models $M0$ and $M2$, and $M0$ and $M1$. [See table \ref{Tab:info_Tranf_formula}]. The analytical expression for the $n-$step KL rate will be used to make the case for discovering a  comparison metric for a model reduction based on the finite time horizon analysis. We will make this case under the simplified setting as stated in the following assumptions.

\begin{assumption}\label{Assume1}
\[\alpha_0 < \beta_0 < \beta_\infty < \alpha_\infty\]
where, $\alpha_n=\Delta H_n(M0,M1)$ and $\beta_n = \Delta H_n(M0,M2)$.
\end{assumption}

This assumption essentially means that the initial KL rate metric for $M1$ is less than that of $M2$, whereas the asymptotically KL rate metric for $M1$ converges to a higher value than that of $M2$. Due to monotonicity of the n-Step KL rate metric, this assumption asserts that there is a time $n=\bar{n}$ when the two curves of $\alpha_n$ and $\beta_n$ intersect each other.
\begin{assumption}\label{Assume2}
For the sake of simplicity we will assume that $\Sigma_0=I_2$ and $\sigma =1$. 
\end{assumption}
This assumption does not constrain the problem. The treatments shown in this paper could be extended to more general cases, but this assumption is made only to improve readability.
For models $M0$, $M1$ and $M2$, under assumptions \ref{Assume1} and \ref{Assume2}, there exists a time-step $\bar{n}$ such that
\begin{equation}
\begin{array}{lr}
    \alpha_n<\beta_n & \text{if } n< \bar{n} \\
    \alpha_n>\beta_n & \text{if } n> \bar{n}
\end{array}
\end{equation}
Now we make following claim which puts bounds on the time at which the curves of $\alpha_n$ and $\beta_n$ intersect each other.

\begin{claim}
The time step $\bar{n}$ satisfies 
\begin{equation}
    \frac{\ln(\alpha_\infty - \beta_0)-\ln\alpha_\infty}{2\ln A_{11}} <\bar{n}+1 < \frac{\ln(\alpha_\infty - \beta_\infty)-\ln\alpha_\infty}{2\ln A_{11}}
\end{equation}
\end{claim}

\begin{pf}
Consider that the two curves $\alpha_n$ and $\beta_n$ intersect each other at point $c$ at time $\bar{n}$ i.e. $\alpha_{\bar{n}} = c = \beta_{\bar{n}}$. The existence of such point is guaranteed due to Assumption \ref{Assume1}. Hence,
\begin{equation} \label{Eq:Inequality}
\alpha_0 < \beta_0 < c < \beta_\infty < \alpha_\infty    
\end{equation}
Specifically note that $(\alpha_\infty-\beta_\infty)<(\alpha_\infty-c)$ and $(\alpha_\infty-c)<(\alpha_\infty-\beta_0)$. 

At time step $\bar{n}$ equation ~(\ref{Eq:nStepKLSoln}) could be written as 
\[ \alpha_{\bar{n}} = A_{11}^{2\bar{n}} \alpha_0  + (1-A_{11}^{2\bar{n}})\alpha_\infty
\]
Hence, 
\[ 
\bar{n} = \frac{1}{2\ln A_{11}}\ln \left(\frac{c-\alpha_\infty}{\alpha_0-\alpha_\infty} \right)
\]
From the equations for $\alpha_0$ and $\alpha_\infty$ we can write
\[
\alpha_0 = (1-A_{11}^2)\alpha_\infty
\]
Hence, $\bar{n}$ could be simplified to 
\[ 
\bar{n} = \frac{\ln(\alpha_\infty-c)-\ln(\alpha_\infty)}{2\ln A_{11}}-1
\]

From the inequality in equation ~(\ref{Eq:Inequality}) and using monotonicity property of log function $\ln(\alpha_\infty-\beta_\infty)<\ln(\alpha_\infty-c)$ and $\ln(\alpha_\infty-c)<\ln(\alpha_\infty-\beta_0)$.  Also, note that since $\ln(A_{11})<0$ the inequality signs are switched after dividing by $\ln(A_{11})$. 
\end{pf}

\noindent \textbf{Note:} Note that the lower bound on the time of intersection $\bar{n}$ could be made arbitrarily large. When $\alpha_\infty$ and $\beta_0$ are close, $\ln(\alpha_\infty-\beta_0)$ is a large negative value. Also due to the assumption that $\beta_\infty<\alpha_\infty$ we know that the two asymptotic KL rate metrics are different. When $\bar{n}$ is large, it means that $\alpha_n<\beta_n$ for arbitrarily large time, but asymptotically $\alpha_\infty>\beta_\infty$.
In other words, in this case, the asymptotic KL rate metric could not be used to assess the accuracy of models for practical simulation purposes.

\begin{figure}[tb]
\centerline{\includegraphics[scale=1]{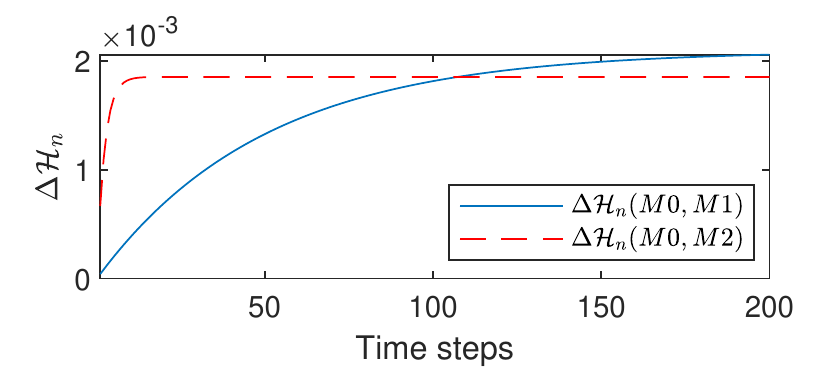}}
\caption{Example of n-step KL-rate metric}
\label{fig:nStepKLRate}
\end{figure}

Fig. \ref{fig:nStepKLRate} shows an example of how comparison between models is dependent on simulation time. Table \ref{tab:Example_vals} shows the specific system parameters used in the example. The lower values of n-step KL rate metric indicate that the model is better approximation. Hence, it can be seen that for first 105 steps, the model $M1$ is better approximation of $M0$, whereas asymptotically $M2$ is a better approximation of $M0$. Hence, for first 105 steps, the model obtained by removing state $a$ is better than the model obtained by removing state $b$. 
\begin{table}[tb]
    \centering
    \captionsetup{width=\linewidth}
     \caption{Values of system parameters used in example}
    \begin{tabular}{|c|c|c|c|c|c|c|c|} \hline
        Parameter & $A_{11}$ & $A_{22}$ & $C_{1}$ & $C_{2}$ & $x_0$ & $\sigma$ & $\Sigma_0$ \\ \hline
        Value & 0.99 & 0.8 & 1 & 0.2 & $\begin{bmatrix}
        1 & 1 \end{bmatrix}^\top$& 1 & $I_{2\times 2}$ \\ \hline
    \end{tabular}
    \label{tab:Example_vals}
\end{table}


Similarly, the Hankel singular values of the model $M0$ reveal that the dynamics corresponding to state $b$ could be removed. In other words, the truncation of states using Hankel singular values would result in the removal of state $b$, but based on the n-Step KL rate metric, for the first 105 steps removing state $a$ provides a better approximation. The point is that for practical purposes, using the n-step KL rate metric may provide better approximation than the traditional asymptotic metrics.

\section{Information Transfer}
{
We demonstrated using simple examples that asymptotic model comparison is not necessarily sufficient for model reduction over a finite time horizon. For practical purposes, n-step KL rate metric provides a comparison over the time span of interest. The overall model reduction process has two steps 
\begin{itemize}
    \item \textbf{Step 1:} Construct simplified models by considering some states to be constants
    \item \textbf{Step 2:} Compute n-step KL rate metric to compare all the simplified models with the original model
\end{itemize}

Motivated by the process of model reduction, we propose a definition of information transfer which essentially combines the concept of freezing of states (i.e. truncation of states) with the n-step KL rate metric. 

}
\subsection{Definition of IT}

\begin{definition} [State-to-State IT]
Consider a linear system $v_{t+1}=Av_t+d_t $, where $v\in\Re^m$ is a state vector, and $A$ is Piccard . The state space $v$ can be divided into two subspaces according to $a\in\Re^{m_a}$ and $b\in\Re^{m_b}$.  The n-step IT from state $a$ to state $b$ for this dynamical system is defined as 
\begin{equation}
    [T_{a \rightarrow b}]_0^{n} = \mathbb{E}_{p(b^{n})}\left[-\ln \left(\frac{p(b_{n}|x_0^{n-1})}{q(b_{n}|x_0^{n-1})} \right)\right]
\end{equation}
where, $q$ is a pdf obtained using an approximate model obtained by considering that state $a$ are constant.

\end{definition}

\begin{definition} [State-to-Output IT] \label{Def:S2OIT}
The $n$ step IT from $a$ to $y$ for dynamical system $\cM0$ in \ref{model1} at time $t$ is denoted by $[T_{b \rightarrow y}]_t^{n}$ and given  by
 \begin{align} 
     [T_{a \rightarrow y}]_0^{n} &= \mathbb{E}_{p(y^{n})}\left[-\ln \left(\frac{p(y_{n}|y_0^{n-1})}{q(z_{n}|y_0^{n-1})} \right)\right] \\
     &= \mathcal{H}_n(\cM0,\cM1)
 \end{align}
 
 \noindent where, $q$ is a pdf obtained using an approximate model $\cM1$ obtained by freezing state $a$ as shown below
 \begin{eqnarray*}
     \begin{bmatrix} a_{t+1} \\ b_{t+1} \end{bmatrix} &=& \begin{bmatrix} 1 & 0 \\ A_{21} & A_{22}\end{bmatrix} \begin{bmatrix} a_t \\ b_t \end{bmatrix} + Bd_t \nonumber \\
    z_t &=& Cw_t
    \label{Eq:LinSys_Frz}
 \end{eqnarray*}
\end{definition}
Although here the definition is based on linear systems, it can be easily extended to nonlinear systems. Here we state some properties of IT. The state-to-state IT is a special case of state-to-output IT.  Also, note here that the IT is a property of the model (and not just a metric of comparing two models). 

\subsection{Properties of IT}
 It can be shown that IT is asymmetric, there exists zero information transfer and information is always conserved. The detailed proofs for slightly different definition of IT are available in \cite{sinha2017information}. 


\subsection{Model Reduction using IT}
Consider that model $\cM0$ is the 'true model'. The two reduced models can be obtained by either considering state $a$ to be constant or state $b$ to be constant. The model with constant $a$ is essentially the model $\cM1$. The model with constant $b$ is
\begin{equation}
\cM2: 
\begin{cases} 
 \begin{bmatrix} a_{t+1} \\ b_{t+1} \end{bmatrix} = \begin{bmatrix} A_{11} & A_{12} \\ 0 & 1\end{bmatrix} \begin{bmatrix} a_t \\ b_t \end{bmatrix} +  Bd_t \nonumber \\
    z_t = Cw_t
\end{cases}
\end{equation}
 
For the model $\cM0$ we make following claim
\begin{claim}
If
\begin{equation}
  [T_{a \rightarrow y}]_0^{n}<[T_{b \rightarrow y}]_0^{n}  
\end{equation}
then the reduced order model obtained by retaining only the state $b$ (i.e. $\cM1$) has following property:
\begin{equation}
    \Delta\mathcal{H}(\cM0,\cM1)<\Delta\mathcal{H}(\cM0,\cM2)
\end{equation}
\end{claim}
The proof of the claim follows from definition \ref{Def:S2OIT}.
This claim allows model reduction using IT as a metric to compute the influence of the states on the outputs. If IT from states $a$ is less than IT from states $b$ then the model could be reduced by truncating states $a$.
\subsection{Numerical Example}
\begin{figure}[tb]
\centerline{\includegraphics[scale=0.85]{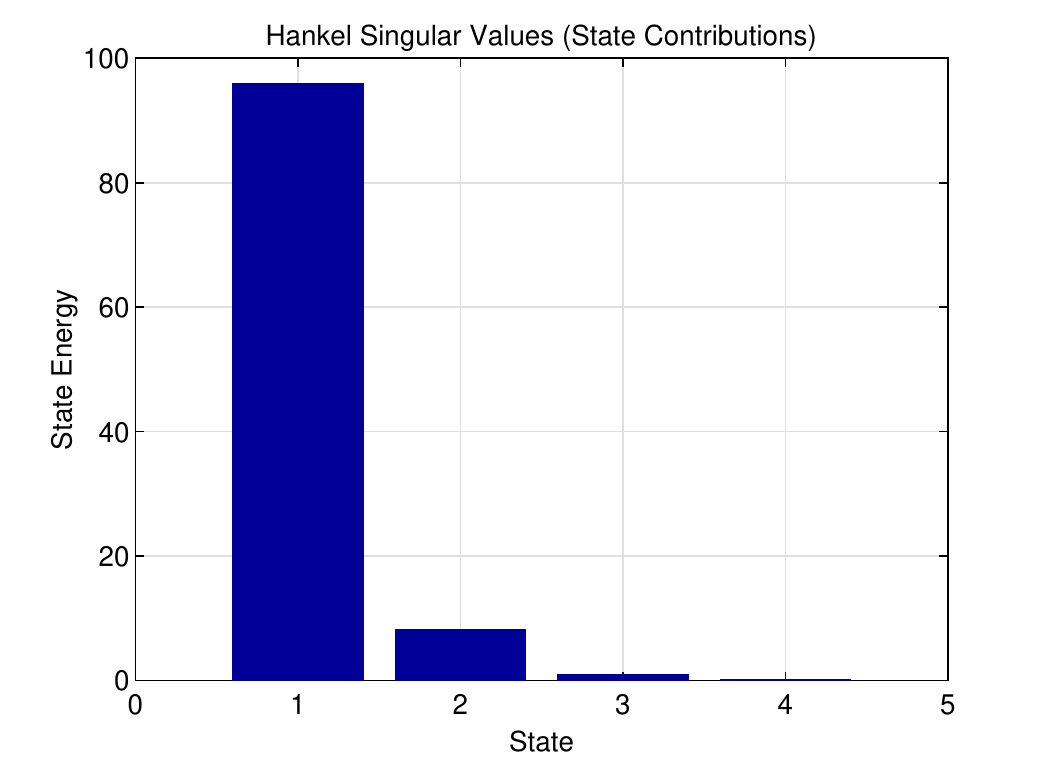}}
\caption{Hankel Singular Values}
\label{fig:HankelSing_2}
\end{figure}

Consider following linear system model. It is already expressed in form of balanced realization. 
\begin{eqnarray} \label{Eq:ModelEx2}
x_{t+1} &=& \begin{bmatrix}  
    0.991  &  0.015  & -0.007 &    0.003 \\
   -0.006  &  0.927 &   0.074  & -0.034 \\
    0.001  & -0.015 &   0.813  &  0.195\\
   -0.000  & -0.002  &  0.025   & 0.309\end{bmatrix} x_t
   +Bd_t \nonumber \\
y_t &=& \begin{bmatrix} 1.281  & -1.065  &  0.506 &  -0.237 \end{bmatrix}x_t 
\end{eqnarray}
Where $B$ is an identity matrix. The objective is to reduce this model by two states. There are 6 combinations of states which could be truncated, for example, the states 3 and 4, or states 1 and 4, etc. Hankel singular values for the model are shown in Fig. \ref{fig:HankelSing_2}. By looking at the Hankel singular values it may be concluded that the states 3 and 4 should be truncated. 

The IT from all six combinations of states to outputs is computed by following the steps below.

\begin{itemize}
    \item A set of combinations of two states is generated as $S = \{(1,\;2),\;(1,\; 3),\; (1,\; 4),\; (2,\;3),\; (2,\;4),\; (3,\;4)\}$.
    \item A set of models with states corresponding to each element in set $S$ held constant are generated. Let's call this set of models as $\cM_i\; \forall \; i=1,...,6$.
    \item n-Step KL rate metric with respect to the original model in equation ~(\ref{Eq:ModelEx2}) is computed for all models i.e. $\Delta\mathcal{H}_n(\cM0,\cM_i)\; \forall i$. This n-step KL rate metric is the IT from states to outputs.
\end{itemize}

\begin{figure}[htbp]
\centerline{\includegraphics[scale=0.7]{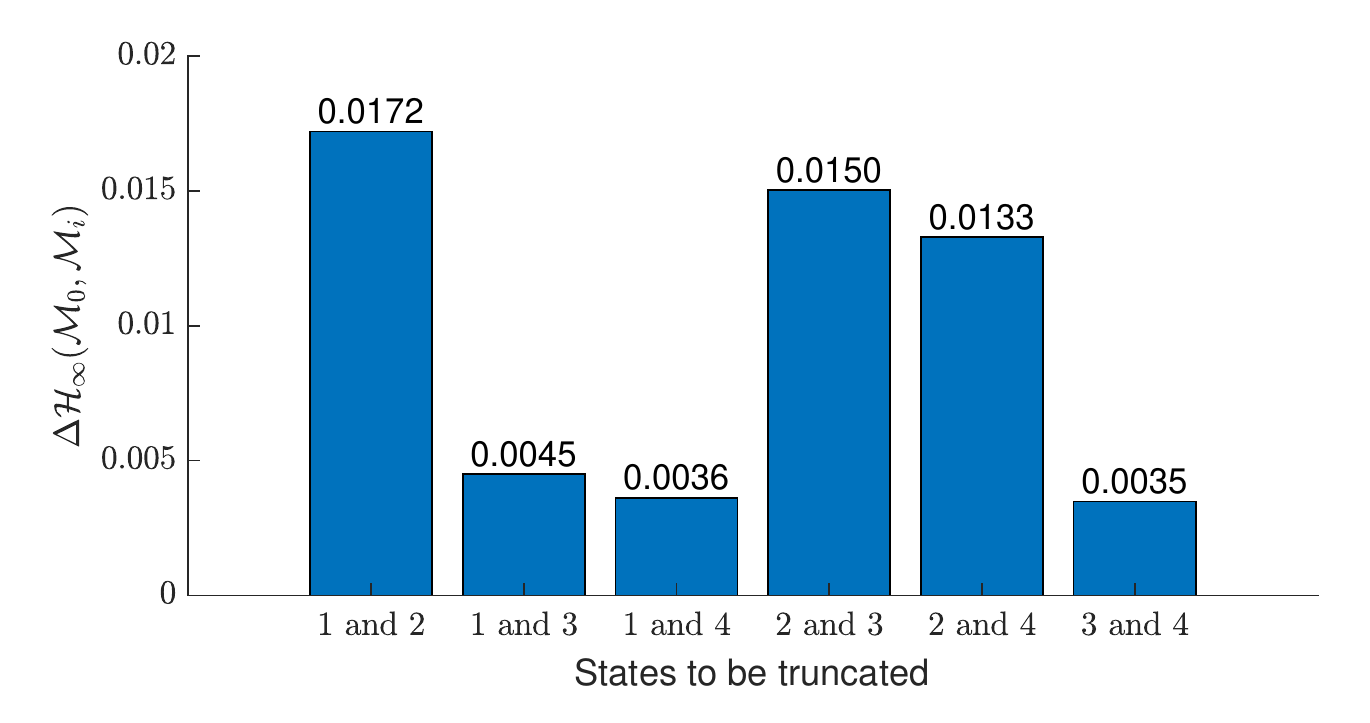}}
\caption{Asymptotic KL rate metric between the original model $cM0$ and all $\cM_i$. Each $\cM_i$ corresponds to truncation of two states from set $S$. Higher values of KL rate metric indicate larger difference between the models.}
\label{fig:dHinf_2}
\end{figure}

\begin{figure}[!tb]
\centerline{\includegraphics[scale=0.7]{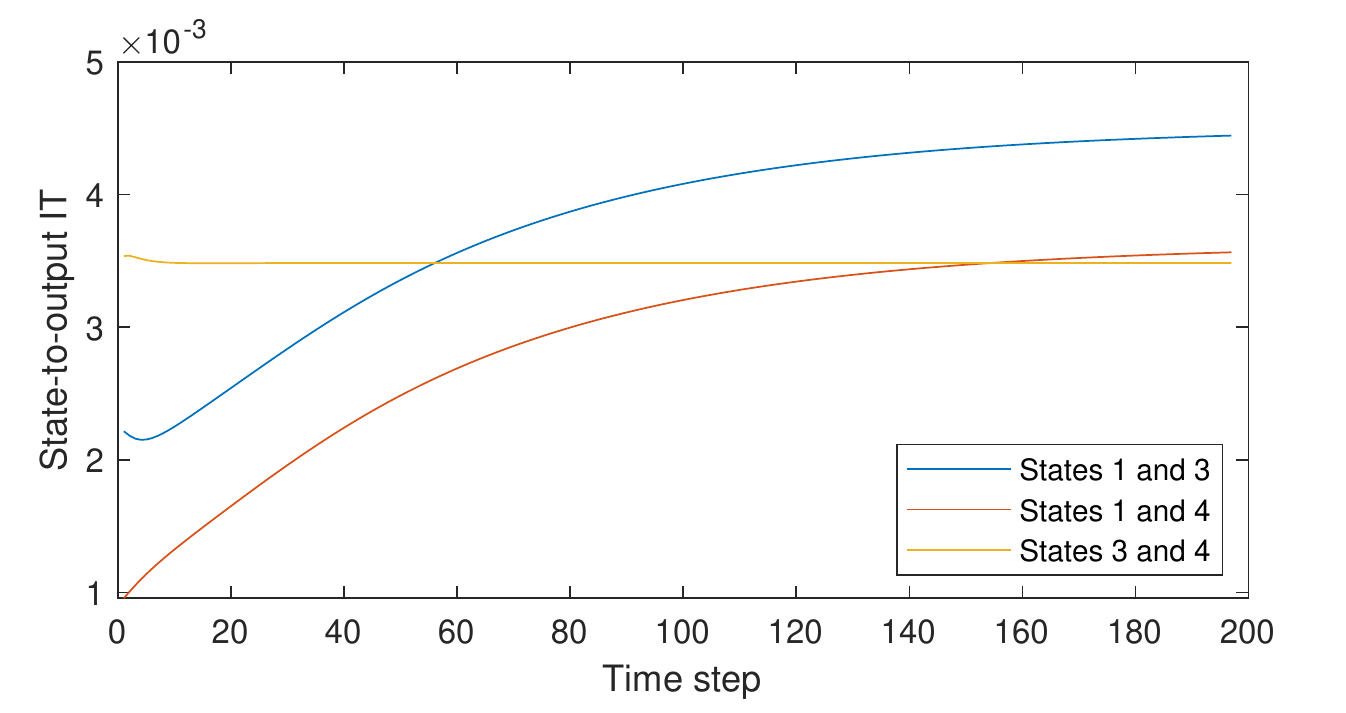}}
\caption{Information transfer from states to outputs}
\label{fig:IT_2}
\end{figure}

Fig. \ref{fig:dHinf_2} shows asymptotic KL rate metric between original model $\cM0$ and all the reduced models $\cM_i$. Each $\cM_i$ represents a model generated by truncating states in the elements of the set $S$. Hence, the bar graph shows the error in prediction due to the removal of certain states. Similar to Hankel singular values, it can be concluded that the removal of states 3 and 4 results in the least error in an asymptotic sense.
The n-step IT for three combinations of states with the smallest IT is shown in Fig. \ref{fig:IT_2}.  It can be observed that the IT from states 1 and 3 to outputs is lower than the IT from state 3 and 4 for some finite time. Similarly, IT from states 1 and 4 is lower for some time, although asymptotically, states 3 and 4 have the lowest IT. The asymptotic assessment of the states to output relation matches the assessment from Hankel singular values, but over finite time other states have smaller influence.  Hence, it could be concluded that the reduced model obtained from removing states 3 and 4 is a better approximation for asymptotic considerations, but the model obtained from removing stats 1 and 4 may be better for practical simulation purposes.

\section{Conclusion}
Metrics and pseudo metrics such as KL rate metric and Hankel singular values could be used to asymptotically compare dynamical system models. Model fidelity and accuracy analysis are required to be able to generate a most accurate model with the least number of states. Hence, the model reduction procedure makes use of these metrics along with state removal or states truncation.  In this paper, we demonstrated using simple examples that these metrics may not be employed for comparing models and model reduction purposes for finite time simulations.  It was shown that a model that is asymptotically better than other, may perform poorly over shorter time scales. Hence, there is a need to develop a generalization of these metrics which could be easily used over shorter time scales. Motivated by this example, we proposed an n-Step KL rate metric which provides a more pragmatic approach of comparing models. Further, we proposed a unique definition of Information Transfer which combines the concept of n-step KL rate metric and removal of states for model reduction. 
 IT could be used to decide which states could be truncated depending upon the time horizon of interest. 
Besides, IT quantifies the influence of states on other states and states on observables. Hence, IT is not just a metric to compare one model against another or a method of model reduction, but it is a property of a model itself. Although this paper is restricted for linear systems, the analysis presented and the proposed IT definition could be readily extended for certain classes of nonlinear systems. IT could also be used for systems with inputs. For such systems, the operating conditions or loading cycles would impact the choice of models.

\bibliography{BibList}
\end{document}